# Type-controlled Nanodevices Based on Encapsulated Few-layer Black Phosphorus for Quantum Transport

*Gen Long[1,2], Shuigang Xu[1,2], Junying Shen[1,2], Jianqiang Hou[1], Zefei Wu[1], Tianyi Han[1], Jiangxiazi Lin[1], Wing Ki Wong[1], Yuan Cai[1], Rolf Lortz[1], Ning Wang[1,3]*

**Affiliations**

[1] Department of Physics and Center for 1D/2D Quantum Materials, The Hong Kong University of Science and Technology, Clear Water Bay, Hong Kong, China

[2] These authors contributed equally.

[3] Correspondence to: phwang@ust.hk

**Abstract**

**We demonstrate that encapsulation of atomically thin black phosphorus (BP) by hexagonal boron nitride (h-BN) sheets is very effective for minimizing the interface impurities induced during fabrication of BP channel material for quantum transport nanodevices. Highly stable BP nanodevices with ultrahigh mobility and controllable types are realized through depositing appropriate metal electrodes after conducting a selective etching to the BP encapsulation structure. Chromium and titanium are suitable metal electrodes for BP channels to control the transition from a p-type unipolar property to ambipolar characteristic because of different work functions. Record-high mobilities of 6000 $cm^2V^{-1}s^{-1}$ and 8400 $cm^2V^{-1}s^{-1}$ are respectively obtained for electrons and holes at cryogenic temperatures. High-mobility BP devices enable the investigation of quantum oscillations with an indistinguishable Zeeman effect in laboratory magnetic field.**

## 1. Introduction

As an emerging two-dimensional (2D) material, atomically thin black phosphorus (BP) has been extensively investigated because of its unique electronic and optical properties[1-4]. High carriers mobilities reaching 1000 $cm^2V^{-1}s^{-1}$ promoted its applications[5]. Unfortunately, the instability of this material in ambient condition degrades quality[6], and thus limited it applications. Recently, encapsulating BP by hexagonal boron nitride (h-BN) sheets has been proved to be very effective for protecting BP from oxidation and largely reducing surface impurity effects[7], and high quality BP nanodevices with hole mobility up to several $10^3$ $cm^2V^{-1}s^{-1}$ has been fabricated based on encapsulated structures[8]. Nevertheless, the full control of the type (p-type and n-type) of high quality BP devices still remains challenging. Fabrication of high-quality low-temperature ohmic contacts for both holes and electrons simultaneously, between BP and metal electrodes, are difficult mainly because of work function mismatches and distinct effective masses of holes and electrons[9]. The Fermi level pining effect[10] originated from impurities at the interfaces which normally cause high Schottky barriers and low injection current should be another obstacle to high quality BP nanodevices. In our study, the well-developed selective etching process to fabricate FE devices based on h-BN/BP/h-BN heterostructures[7] is applied to obtain high-quality ohmic contacts for BP channels. The polarities of BP transistor can be controlled by choosing appropriate contact metals with suitable work functions. We respectively obtain high FE mobilities of 6000 and 8400 $cm^2$ $V^{-1}$ $s^{-1}$ for electrons and holes, with the corresponding Hall mobilities of approximately 3400 and 4800 $cm^2$ $V^{-1}$ $s^{-1}$, in few-layer BP devices. Shubnikov-de Hass (SdH) oscillations are observed clearly in the laboratory magnetic field with an intermediate strength of approximately 6 T. Lifshitz-Kosevich (LK) equation is applied to interpret the quantum transport behavior and to determine relevant parameters, such as Berry phase (*β*) and cyclotron mass in BP 2D electron and hole gases.

## 2. Results and discussion

*2.1 Fabrication of Nanodevices based on h-BN encapsulated BP*

BP is easily oxidized under atmospheric condition[6, 7], encapsulation of atomically thin BP by h-BN sheets is essential for fabricating highly stable BP FE devices (Fig.1a). In our study, the h-BN/BP/h-BN sandwiched structures are prepared in a glove box filled with high-purity nitrogen. A polymer-free transfer technique[11] is applied to ensure ultra-clean BP-BN interfaces. The thickness of the h-BN sheets selected for the sandwiched structures is critical. Because an excessively thin h-BN is unable to screen the charge impurity scattering effects[12, 13] from the $SiO_2$ substrates used to support the FE devices[14]. Therefore, h-BN flakes thicker than 10 layers or stacked together with thick graphene flakes are selected as the substrates for our devices. The top BN placed on the BP channel effectively protects BP from oxidation (Supplementary Materials). The results show that a high-temperature annealing (250~350 °C for 15 hrs) is beneficial for stabilizing the BP devices. Before metal electrodes are deposited, a Hall device configuration is defined through a standard electron beam (e-beam) lithography technique. To expose BP surfaces, Selective etching based on oxygen plasma is performed to expose BP surface. The etching rate of oxygen plasma for BN is much faster than that for BP. Thus, the top BN is etched away while the BP flakes survive. Then Ti/Au electrodes are deposited by employing a standard e-beam evaporation technique (Fig. S1). The thicknesses of BP flakes are initially estimated according to their contrasts and then precisely measured by an atomic force microscope (AFM) (Fig. S2). Angle-resolved Raman spectroscopy is conducted to determine the crystallographic direction of the BP flakes (Fig. S3)[15]. The BP channels in our devices are placed along the X direction, and these channels exhibit higher mobility than those along the Y direction[16].

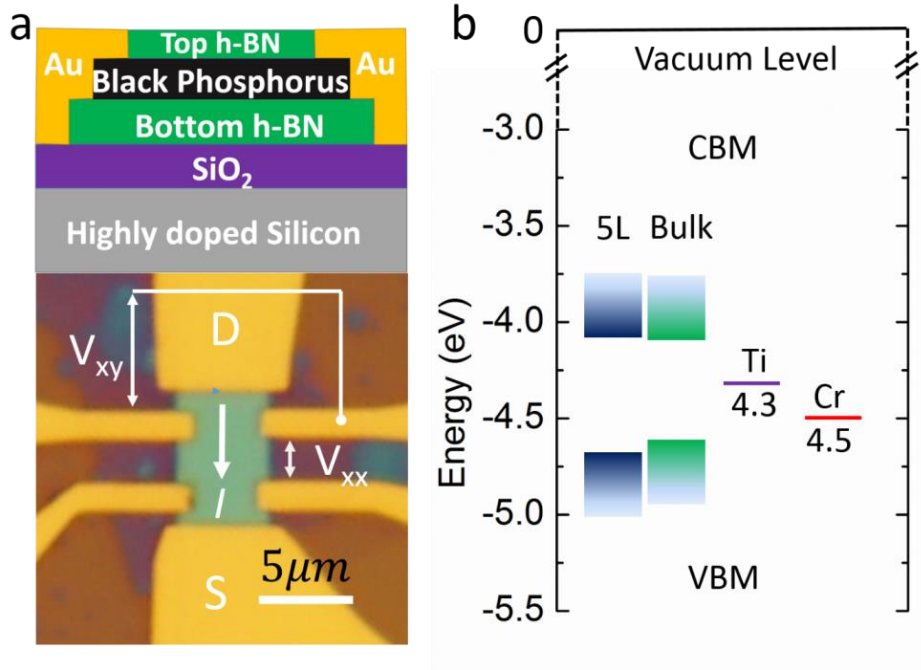

**Fig. 1. The structure of a BP FET.** (**a**) The top panel shows the layer structure of h-BN/BP/h-BN heterostructures. The bottom panel is an optical micrograph showing a typical BP device as well as the measurement configuration. The scale bar is 5 $\mu m$. (**b**) Work functions of chromium, titanium and BP.

*2.2 Polarity control of BP based field effect transistors (FETs)*

The band alignment conditions at the interface between the BP channel and metal electrodes are critical for the polarities of quantum transport in BP devices[17]. As shown in fig.1b, Cr has a work function aligned well with the valence band maximum while the work function of Ti falls in the gap of BP[18]. Consequentially, BPs contacted with Cr serve as high-performance p-type channels; and those contacted with Ti function as ambipolar transistor channels. The right panel in Fig. 2c shows that, the current ($I_{ds}$) injected into sample A with Cr contacts depends linearly on excitation voltages ($V_{ds}$) under different gate voltages (or carrier concentrations) at cryogenic temperatures. This result demonstrates a high-quality ohmic contact between the BP flakes and the metal electrodes. The transport characteristics (Fig. 2a) of the BP device at cryogenic temperatures also display typical *p*-type unipolar features. The p-type polarity of our BP devices contacted with Cr electrodes is consistent with previous reports[7, 19-21]. The left and middle panels of Fig.1c reveal the $I_{ds}$-$V_{ds}$ characteristics of sample B with Ti contacts. The $I_{ds}$ depends linearly on $V_{ds}$ on both hole and electron sides. Figure 2a illustrates the characteristics of an ambipolar transistor channel.

*2.3 Record-high mobilities of holes and electrons in BP devices*

The temperature dependences of the field effect mobility $\mu_{FET}$ and Hall mobility $\mu_H$ evaluated from samples A and B with different polarities are shown in fig.1b. The room temperature $\mu_{FET}$ of sample A (p-type unipolar) and sample B (n-type ambipolar) reach 460 cm$^2$V$^{-1}$s$^{-1}$ and 350 cm$^2$V$^{-1}$s$^{-1}$, respectively, and these values are comparable to that of the widely used silicon. As the devices cool down to 2K, $\mu_{FET}$ increases by almost 20 times to record-high values of 8400 and 6000 cm$^2$V$^{-1}$s$^{-1}$ for the holes and electrons of samples A and B, respectively, and these values are comparable to currently recorded values[8]. The Hall mobility $\mu_H$ shows similar temperature dependence with $\mu_{FET}$. At temperatures lower than 30K, both $\mu_{FET}$ and $\mu_H$ weakly depend on temperature, indicating that impurities and/or disorder scattering dominate over phonon scattering[12, 13, 22]. Therefore, the carrier mobility does not increase by further decreasing the sample temperature in this temperature regime[22]. At temperatures above 100K, the measured mobility can be fitted as $\mu \sim T^{-\gamma}$ (phonon scattering effects), with $\gamma$=1.6~1.8. The values of γ in our samples are larger than that previously measured in BP[23]. We believe that impurity effects and contact problems seriously influence the detection of γ and intrinsic mobility. A large value of γ implies that acoustic phonon related interactions instead of optical phonon related interactions dominate over residual impurity related interactions in this temperature regime.

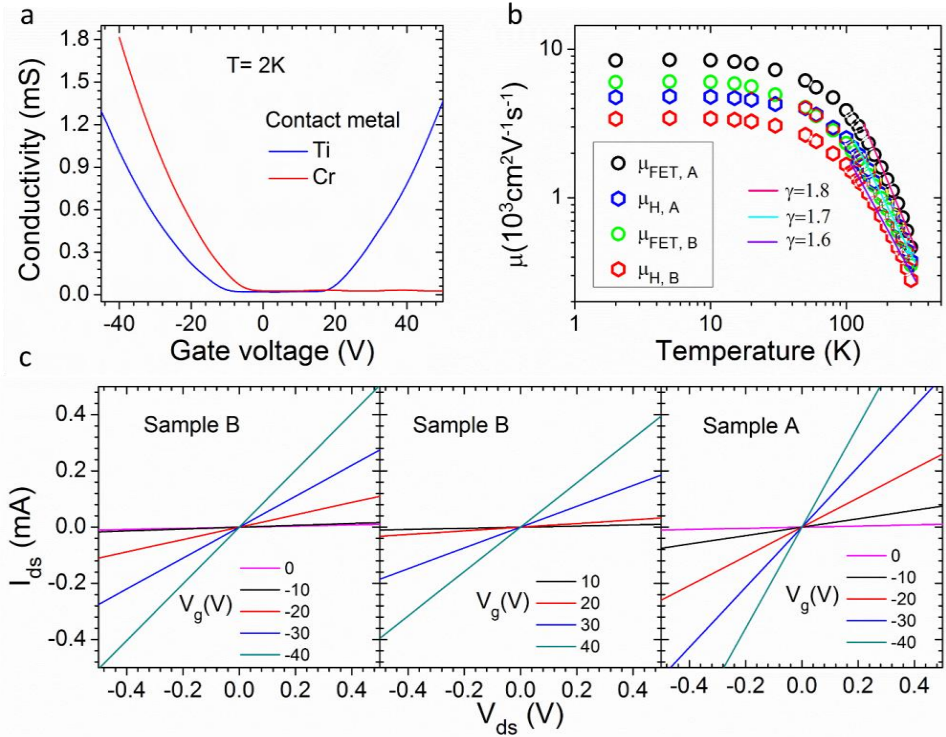

**Fig. 2. Transport properties of BP FETs.** (**a**) Transport curves obtained from four-terminal devices with different contact metals at cryogenic temperatures. (**b**) The FET mobility and Hall mobility for devices with different contact metals measured at various temperatures. The linear fittings are applied to extract γ in the phonon limited region (100K~300K). (**c**) $I_{ds}$-$V_{ds}$ curves obtained by two-terminal configuration at cryogenic temperature. Right panel: p-type sample B with Ti contacts; Middle panel: n-type sample B with Ti contacts; Left panel: p-type sample A with Cr contacts.

*2.4 Quantum Oscillations in BP devices*

SdH oscillations in BP devices can be observed on the basis of high carrier mobility at cryogenic temperatures in laboratory magnetic fields. SdH oscillations in longitudinal resistivity provides important information for accessing certain aspects of 2D electron gas (2DEG) behavior in few-layer BP. Fig. 3a shows the magnetic field (B) dependence of channel resistances for different types of polarities. Pronounced oscillations for both n-type (left panel) and p-type (right panel) BPs are clearly visible in the magneto-resistance plots, and the filling factors are indicated by green arrows. The quantum oscillations originate from Landau quantization of the cyclotron motion of the charge carriers in BP, which show a typical 2D nature as revealed in a previous work[19]. The quantum oscillations With an indistinguishable Zeeman effect under intermediete strengths of magnetic field, can be described with LK formula[24], expressed as follows:

$$\Delta R = R(B,T)\cos(2\pi(B_F/B + 1/2 + \beta)) \qquad (1)$$

where $R(B,T)$ denotes the oscillation amplitude and $\beta$ is the Berry phase. The oscillation periods $1/B_F$ of the channel resistance ($R_{xx}$ vs. $1/B$) at different gate voltages can be obtained through fast Fourier transform (FFT) analyses (insets of fig.3a). The carrier concentration can be calculated on the basis of oscillation periods by considering the charge carrier density $n = g_s e B_F / h$, where $g_s = 2$ denotes the spin degeneracy. Changes in the carrier concentrations obtained from the oscillations at different gate voltages are presented in fig. 3b (hollow circles) and the solid line in the same figure represents the linear fitting results based on a capacitance model. A capacitance of $7 \times 10^{10} ecm^{-2}V^{-1}$ is obtained from the fitting results. The charge carrier concentrations are also obtained on the basis of Hall effects. The results of oscillation periods are consistent with those of Hall effects. The inset in Fig.3b revealed the enlarged fitting results in the low carrier concentration regime. A band gap of about 0.57 eV is obtained, and this result is consistent with the calculated band structure of a few layer BP[9].

Landau level ($N$), where $N = B_F / B$ and $B$ is the corresponding magnetic field when the chemical potential is at the middle of two adjacent Landau levels, should be identified to determine $\beta$ in our 2D BP samples. Fan diagrams of $N$ and its linear fittings with respect to $1/B$ are shown in Fig. 3c. The linear fitting yields $\beta$ of 0 for both p-type and n-type BP 2D carrier gases. A non-zero β of $\pi$ cannot be obtained from our data, which are different from previously reported results[19], in which quantum oscillations and distinguishable Zeeman effects have been observed synchronously in a high magnetic field of approximately 20 T. However, the LK model may be invalid for SdH oscillations with a distinguishable Zeeman splitting in high strengths of magnetic field[25]. The slopes of the fan diagrams in a low magnetic field with an indistinguishable Zeeman effect may differ from those in a high magnetic field with distinguishable Zeeman effect[26]. This difference may lead to an artificial $\beta$ calculated from the fan diagrams in a high magnetic field. At a phenomenological level when Zeeman effects are considered, the LK formalism can be expressed as follows:

$$\Delta R = \sum_{\uparrow,\downarrow} R(B,T) \cos(\phi_{\uparrow,\downarrow}) \qquad (2)$$

where the phase $\phi_{\uparrow,\downarrow} = 2\pi(B_F / B)\{1 \mp [(g\mu_B B)/(2\hbar\omega B_F / B)]\} - \pi$ contains the Zeeman component $g\mu_B B$ [27, 28]. Under a magnetic field of weak strength, the Zeeman Splitting energy $g\mu_B B$ is much smaller than the Fermi energy $\hbar\omega B_F / B$, and eq. (2) can be converted to eq. (1). For strong magnetic fields, $g\mu_B B$ is comparable to $\hbar\omega B_F / B$, thus a non-zero artificial berry phase should come from the term $\frac{g\mu_B B}{2\hbar\omega B_F / B}$. In contrast to previously reported data, the $\beta$ determination in the present study is based on the oscillations measured with an indistinguishable Zeeman splitting. And a zero $\beta$ is extracted from the fan diagram. Fig. 3d illustrates the relative energy spacing of the cyclotron energy $E_c = \hbar eB/m^*$ and the Zeeman splitting energy $E_z = g\mu_B B$. The dips in SdH oscillations of our samples correspond to quantum states with even filling factors for both kinds of polarities. This phenomenon could be ascribed to $E_z / E_c < 0.5$. Thus, spin susceptibility $\chi_s = gm^* < 1$. Here, $m^*$ is the cyclotron mass in the unit of $m_0$ and $g$ represents the Lande g-factor.

The cyclotron mass of the carriers in our BP samples can be determiend experimentally on the basis of the temperature dependence of magneto-resistance oscillation amplitudes (Fig. 3e) $\Delta R_T \propto \lambda(T)/\sinh(\lambda(T))$, where $\lambda(T) = 2\pi^2 k_B T m^* / \hbar eB$ is the parameterized thermal damping, $k_B$ is the Boltzmann constant, $\hbar$ is the reduced Plank constant and $m^*$ is the cyclotron mass. Fig. 3f illustrates the observed thermal damping at different magnetic fields (hollow dots) and the solid lines represent the fitting results obtained through the LK formula. As shown in Fig. 3g, the average cyclotron masses obtained from the fitting results are 0.32$m_0$ and 0.26$m_0$ for electrons and holes, respectively. The measured cyclotron masses for holes is consistent with the following reported values: 0.26-0.31$m_0$[21], 0.24±0.02$m_0$[29] and 0.27$m_0$[7]. These values are consistent with those obtained from spectroscopic measurements[30]. Considering that $\chi_s < 1$, we determine that the upper bounds of the Lande g-factors for holes and electrons are 3.1 and 3.8, respectively.

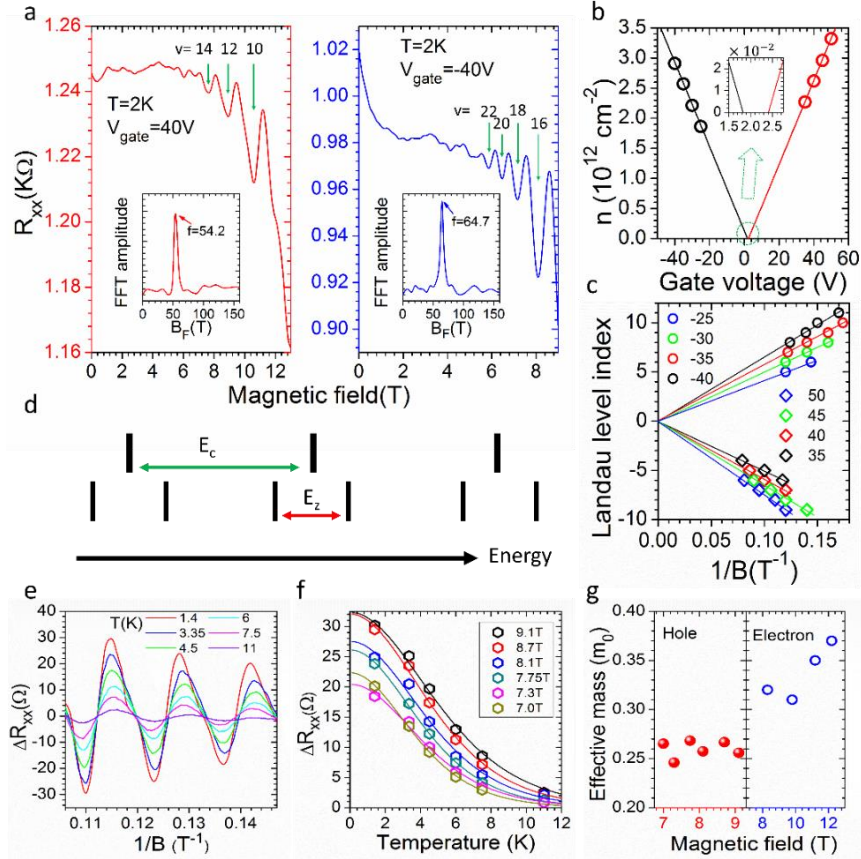

**Fig.3. Shubnikov-de Hass oscillations from ambipolar BP FETs.** (**a**) The longitudinal resistance ($R_{xx}$) of sample B plotted as a function of magnetic field. Right panel: n-type; Left panel: p-type. The insets show the FFT results of $R_{xx}$-1/B curves. (**b**) Carrier concentration obtained from $B_F$ which is extracted from a. (**c**) Landau index fan diagrams for the SdH oscillations at different gate voltages. Landau index for holes and electrons are shown with positive and negative indices for clarity. (**d**) Schematic of the Landau level arrangement with Zeeman splitting. (**e**) 1/B dependence of SdH oscillation amplitudes ($\Delta R_{xx}$) at different temperatures with $V_g$ fixed at -50V. (**f**) Temperature dependence of SdH oscillation amplitudes ($\Delta R_{xx}$) at different magnetic fields. Solid lines are fitting results according to the LK formula. (**g**) The cyclotron mass values obtained from fitting the SdH oscillation amplitudes at different magnetic fields. Right panel: hole; Left panel: electron.

## 3. Conclusion

In summary, the overall quality of atomically thin BP FE devices is significantly improved by encapsulating a BP channel material with h-BN sheets. In particular, type-controlled BP FE devices with high-quality ohmic contacts are obtained by depositing appropriate metal electrodes after conducting a selective etching to the BP encapsulation structure. SdH oscillations in ambipolar few-layer BP can be observed on the basis of high carrier mobility. The LK formalism is applied to interpret transport behavior and to determine quantum transport parameters.

## 4. Methods

Bulk phosphorus crystals are purchased from Smart-elements and BN crystals (Polartherm grade PT110) are purchased from Momentive. The atomically thin flakes of BN and BP are mechanically exfoliated from bulk crystals through the widely used scotch-tape micro cleavage method. Few-layer BP sheets are exfoliated on $SiO_2$/Si supports and h-BN sheets with appropriate thickness (~10 layers) are exfoliated on glass-supported PMMA films which serve as superstrates. Another h-BN sheet prepared on $SiO_2$/Si support will serve as subtract. The superstrate h-BN sheets are used to pick up BP flakes through van der Waals force. The formed h-BN/BP stacks are then put on the prepared subtract h-BN to form h-BN/BP/h-BN sandwich structures. All of these processes are performed in an inert environment of nitrogen to reduce the adsorption of water from air and the oxidation effect. PMMA films are removed by steeping in acetone for 10 min and then isopropyl alcohol (IPA) is applied to flush the residual acetone away. Annealing under $Ar/H_2$ at 300 $^o$C is applied to further stabilize the heterostructure.

Standard electron beam lithography (EBL) technique is then applied to pattern the sandwich structures followed by reaction-ion etching (RIE) processes to define the Hall structures (recite: 4 sccm $O_2$ + 40 sccm $CHF_3$; RF power: 200W). Then we use EBL technique to pattern the contact areas of the defined Hall structures, and the superstrate h-BN is etched by RIE while the BP sheets survive. This can be realized by choosing appropriate recipe which etches h-BN faster than etching BP. Our recipe is $O_2$ 40 sccm and the RF power is 200 W. A small overlap between metal electrodes and superstrate h-BN is induced by the third EBL to make sure all the exposed BP areas are covered by metal electrode. Then standard electron beam evaporations of Ti/Au (5nm/80nm) for ambipolar and Cr/Au (5nm/80nm) for p-type transistors are carrier out to deposited electrode metals on the exposed area.

Electrical measurements are performed using standard lock-in techniques in a cryogenic system (1.5–300 K and magnetic field up to 13.5 T) under low pressure ($3*10^{-3}$ Torr).

**Acknowledgements**

Financial support from the Research Grants Council of Hong Kong (Project Nos. 16302215, HKU9/CRF/13G, 604112 and N_HKUST613/12) and technical support of the Raith-HKUST Nanotechnology Laboratory for the electron-beam lithography facility at MCPF (Project No. SEG_HKUST08) are hereby acknowledged.

**Competing financial interests**

The authors declare no competing financial interests.

**Author contributions**

N. Wang and G. Long conceived the project. G. Long fabricated the devices and performed cryogenic measurements with the help of J. Y. Shen and S. G. Xu. G. Long and N. Wang analyzed the data, and wrote the manuscript. Other authors provided technical assistance in the project.

**Supplementary Materials:**

**Type-controlled Nanodevices Based on Encapsulated Few-layer Black Phosphorus for Quantum Transport**

*G. Long, S. G. Xu, J. Y. Shen, J. Q. Hou, Z. F. Wu, T. Y. Han, J. X. Z. Lin, W. K. Wong, Y. Cai, R. Lortz, N. Wang*

Correspondence to: phwang@ust.hk

**Contents**

1. **Device fabrication**
2. **Determination of sample thickness by AFM**
3. **Determination of the crystallographic direction of BP by Angle resolved Raman spectroscopy**

1. **Device fabrication**

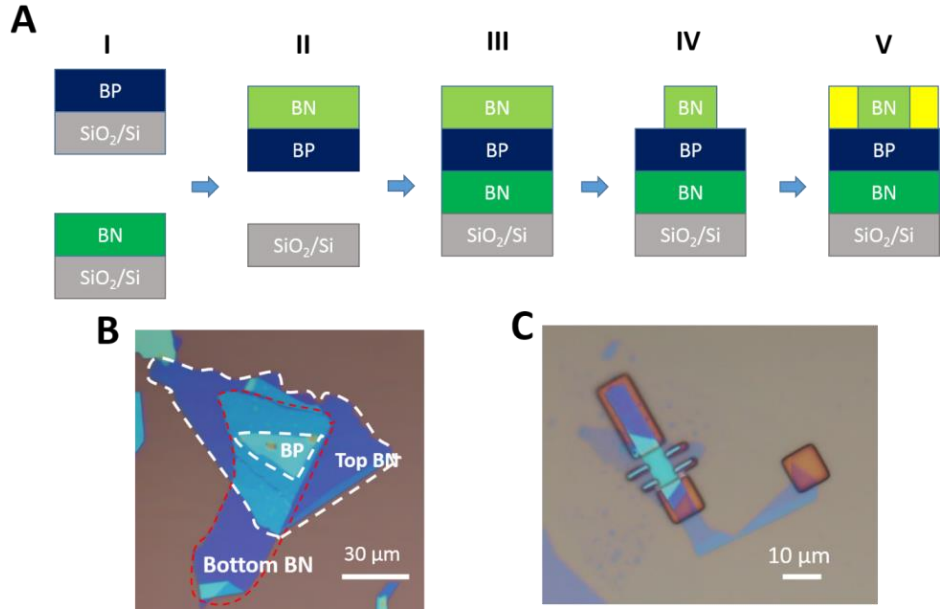

**Fig. S1.** Device fabrication processes. (**A**) Schematic of device fabrication process. (**B**) The optical image of BN/BP/BN hetero-structure after processing by step (III). The BN flake is encapsulated between top and bottom BN. The scale bar is 30 μm. (**C**) The optical image of BP device after the selective etching process. The top BN is etched while the BN reserves. The scale bar is 10 μm.

**Step. I**   The bottom BN and BP flakes are mechanically exfoliated on two separate SiO$_2$/Si wafers. Another BN (top BN) on PMMA film is also prepared.

**Step. II**   The BP is then picked up from SiO$_2$/Si wafer by the top BN.

**Step. III**   The BN/BP structure in Step.2 is transferred to the bottom BN prepared in Step. I.

**Step. IV**   The top BN is etched through the selective etching process.

**Step. V**   Electrode metal is deposited using standard electron beam evaporation technique.

**2. Determination of sample thickness by AFM**

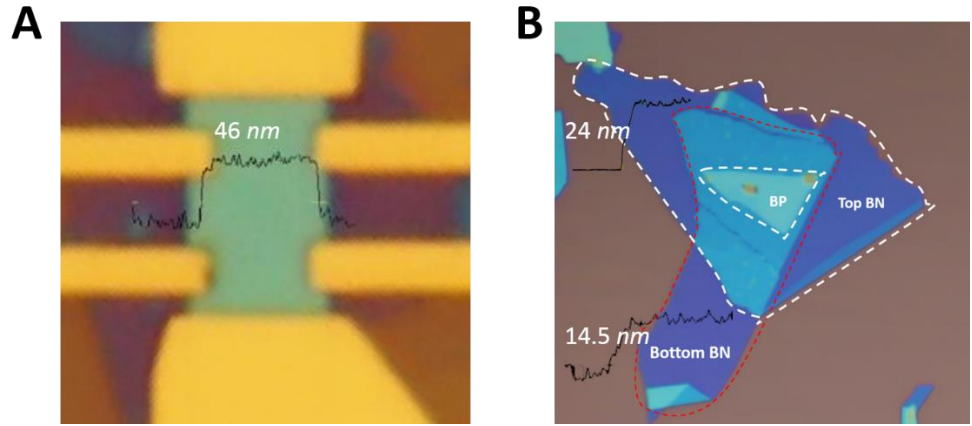

**Fig. S2.** Measurement of the thickness of the BP device using AFM. (**A**) The total thickness of BN/BP/BN hetero-structure is 46 *nm*. (**B**) The thickness of top and bottom BN flakes are 24 and 14.5 *nm* respectively. After calculation, the thickness of BP flake is 7.5 *nm*.

3. **Determination of the crystallographic direction of BP by Angle resolved Raman spectroscopy**

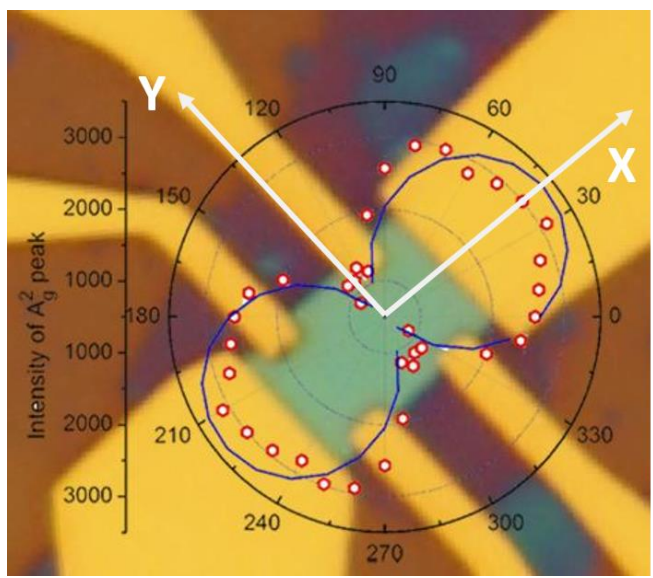

**Fig. S3.** Determination of the crystallographic direction of the BP sample by angle resolved Raman spectroscopy. The polar diagram shows the intensity of $A_g^2$ peak when the polarization direction of the incident laser is in different angles (red dots). The blue line represents the simulation results. The $A_g^2$ peak intensity of BP is maximal (minimal) when the laser polarization is in the X (Y) direction of BP[1]. The crystallographic direction is determined according to the variation of the intensity of $A_g^2$ peak. The crystallographic directions of our sample are indicated by two white arrows marked with X and Y.

**Supplementary References:**